\documentclass[twocolumn,prd,floatfix,nofootinbib]{revtex4}

\usepackage{amsmath}
\usepackage{graphicx}
\usepackage{subfigure}
\usepackage{multirow}
\usepackage{bm}
\usepackage{amssymb}
\usepackage{mathtools}
\usepackage{enumerate}
\usepackage{color}
\usepackage[colorlinks,linkcolor=magenta,anchorcolor=cyan,citecolor=blue]{hyperref}

\begin{document}

\title{Is the Hubble tension a hint of AdS phase around recombination?}

\author{Gen Ye$^{1}$\footnote{yegen14@mails.ucas.ac.cn}}
\author{Yun-Song Piao$^{1,2}$\footnote{yspiao@ucas.ac.cn}}

\affiliation{$^1$ School of Physics, University of Chinese Academy of
    Sciences, Beijing 100049, China}

\affiliation{$^2$ Institute of Theoretical Physics, Chinese
    Academy of Sciences, P.O. Box 2735, Beijing 100190, China}

\begin{abstract}

Anti-de Sitter (AdS) vacua, being theoretically important, might
have an unexpected impact on the observable universe. We find that in
early dark energy (EDE) scenarios the existence of AdS vacua
around recombination can effectively lift the CMB-inferred $H_0$
value. As an example, we study a phenomenological EDE model with
an AdS phase starting at the redshift $z\sim2000$ and ending
shortly after recombination (hereafter the universe will settle
down in a $\Lambda>0$ phase until now), and obtain a best-fit
$H_0=72.74$ km/s/Mpc without degrading the CMB fit compared with
the standard $\Lambda$CDM model.


\end{abstract}

\maketitle

Recently, the tension between the Hubble constant $H_0$ measured
locally and that deduced from the best-fit of $\Lambda$CDM to the
cosmic microwave background (CMB) observation has acquired extensive
attention, e.g.\cite{Verde:2019ivm,Knox:2019rjx} for reviews.
Based on the $\Lambda$CDM model, the Planck collaboration inferred
$H_0=67.36\pm0.54$ km/s/Mpc \cite{Aghanim:2018eyx}. Using
Cepheids-calibrated supernovae, Riess et.al (the SH0ES team)
reported the Hubble rate $H_0=74.03\pm1.42$ km/s/Mpc
\cite{Riess:2019cxk}, which is at $4.4\sigma$ discrepancy compared
with that inferred by Planck. The large $H_0(>70)$ value is also
supported by other local measurements
\cite{Chen:2019ejq,Freedman:2019jwv,Wong:2019kwg,Huang:2019yhh}.
Currently, it is probably not suitable to simply explain this
discrepancy by systematic errors in the data
\cite{Verde:2019ivm}. It is thus increasingly likely that new physics beyond the $\Lambda$CDM model plays a role in resolution of the Hubble tension.

Theoretically, one possibility is modifying post-recombination
physics, such as the dark energy or modified gravity models, e.g.
\cite{DiValentino:2017iww,DiValentino:2017zyq,Raveri:2019mxg,Desmond:2019ygn,Yan:2019gbw,Ding:2019mmw,DiValentino:2019ffd,Visinelli:2019qqu,Akarsu:2019hmw,Colgain:2018wgk,Colgain:2019joh,Li:2019ypi}.
Such solutions are constrained tightly by late-time observations
\cite{Verde:2019ivm}. Another possibility is modifying
prerecombination physics (modifying the sound horizon
$r^*_s=\int_{z^*}^\infty c_s/H(z) dz$
\cite{Verde:2019ivm,Evslin:2017qdn,Aylor:2018drw}), such as early
dark energy
\cite{Poulin:2018cxd,Agrawal:2019lmo,Alexander:2019rsc,Lin:2019qug,Smith:2019ihp,Sakstein:2019fmf,Niedermann:2019olb},
see also \cite{Doran:2006kp},  neutrino self-interaction
\cite{Kreisch:2019yzn,Park:2019ibn}, see also
\cite{Escudero:2019gvw,Blinov:2019gcj,Vagnozzi:2019ezj}, and
decaying dark matter
\cite{Agrawal:2019dlm,Vattis:2019efj,Pandey:2019plg}. See also
attempts concerning CMB non-Gaussianities \cite{Adhikari:2019fvb}
and fundamental constants \cite{Hart:2019dxi}. Determination of
$H_0$ requires fitting the integral expression
$D_A^*=\int^{z_*}_0\frac{dz}{H(z)}=r_s^*/\theta_s^*$, $D_A^*$ being
the angular diameter distance to the last-scattering surface.
While $\theta^*_s\equiv r_s^*/D_A^*$ is precisely determined by
CMB peak spacing, a smaller $r^*_s$ will eventually lead to a
larger $H_0$.

It is well known that anti-de Sitter (AdS) vacua are theoretically important. AdS vacua naturally
emerge from the string theory. One might uplift AdS to de Sitter (dS) vacua by the KKLT
mechanism \cite{Kachru:2003aw,Kallosh:2018psh}, which inspired the
``landscape" idea \cite{Susskind:2003kw}. The landscape consists
of all effective field theories (EFTs) with consistent
UV completion (otherwise the EFT is said to be in the swampland), which might be from various compactifications of the
string theory. As the swampland criteria for EFTs, the distance
conjecture \cite{Ooguri:2006in} and the dS conjecture
$M_p|\nabla_{\phi}V|/V>c\sim {\cal O}(1)$ \cite{Obied:2018sgi} (or
the refined dS conjecture \cite{Ooguri:2018wrx,Garg:2018reu}) have
been proposed, which seems to throw dS vacua into the
swampland. However, whether metastable dS vacua exists
in the landscape or not, AdS vacua should be indispensable. Thus
it is significant to ask if AdS vacua have any impact on the
observable universe.

We will show this possibility. A novelty of our result is that
recombination might happen in the AdS vacuum\footnote{The possibility of late-time AdS has also been studied in Ref.\cite{Akarsu:2019hmw,Visinelli:2019qqu}.}, which may be tested
by near-future CMB experiments. In the early dark energy (EDE)
scenario \cite{Poulin:2018cxd,Agrawal:2019lmo,Smith:2019ihp},
modification of $r_s$ is implemented by an EDE scalar field that
starts to activate a few decades before recombination. It is the
energy injection of this EDE field that results in a reduced
$r^*_s$, so an increased $H_0$. We will focus on the EDE scenario
with an AdS phase, and find that such an AdS phase will make the
EDE injection more efficient while ensuring that it redshifts fast
enough around recombination without spoiling the fit to the CMB data.


\begin{figure}
    \includegraphics[width=\linewidth]{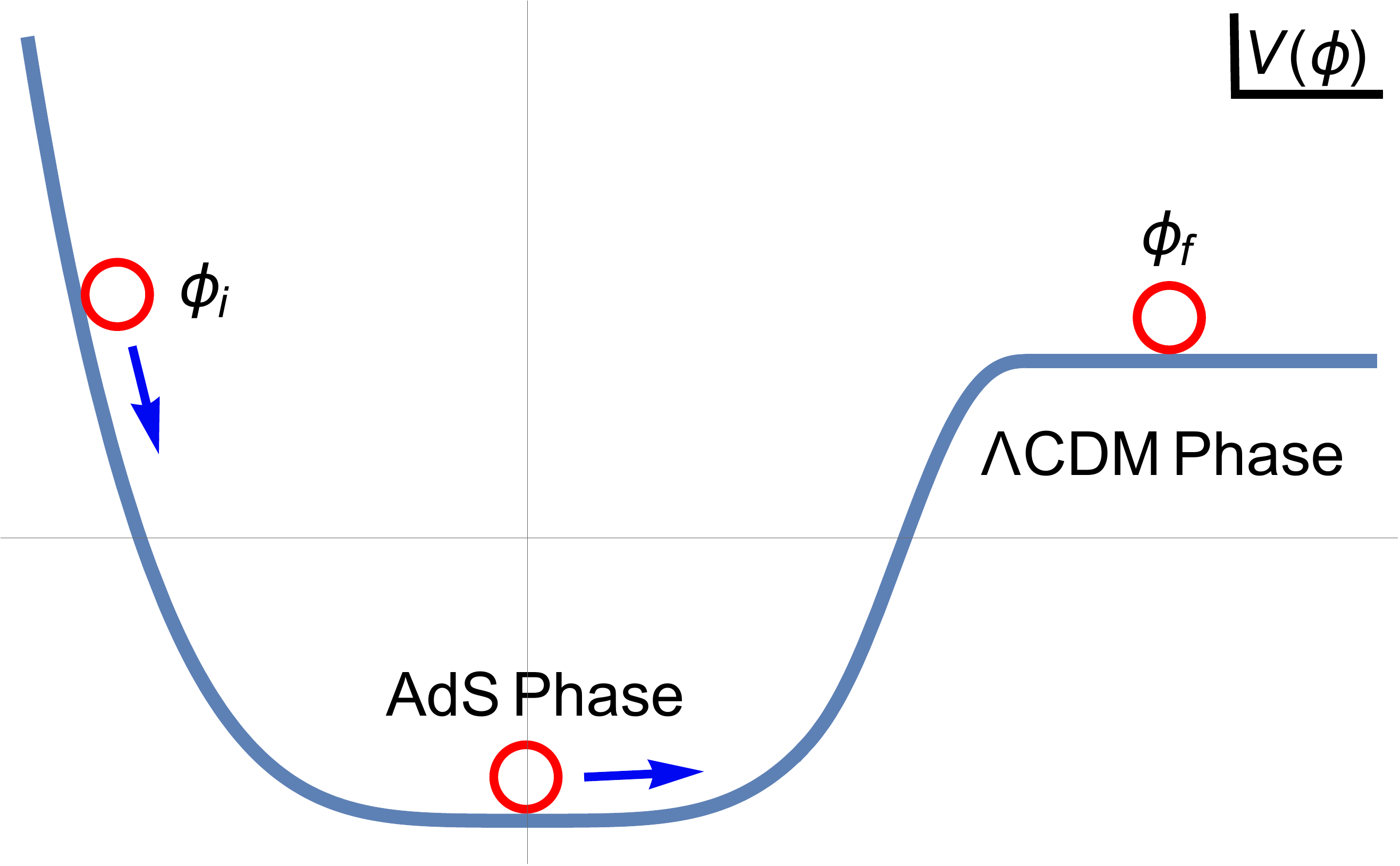}
\caption{A sketch of the potential with an AdS phase. Initially
the field is frozen at $\phi_i$. It starts rolling down the
potential at the redshift $z\sim3500$ when $H$ drops below its
effective mass $m_\phi$, and enters an AdS phase at $z\sim2000$.
The field straightly rolls over the AdS region and does not
oscillate. It climbs up to the $\Lambda>0$ region shortly after
recombination, hereafter the universe is effectively described by
the standard $\Lambda$CDM model.}
    \label{Vads}
\end{figure}

The scenario we consider is presented in Fig-\ref{Vads}.
Initially, the scalar field $\phi$ sits at the hillside of its
potential, and its energy density $\rho_\phi$ is negligible. As
the universe expands, radiation and matter are diluted. When
$H^2\simeq
\partial_\phi^2V$, which occurred before recombination, the field
starts rolling down the potential, meanwhile $\rho_\phi$ becomes
non-negligible. Then the field will roll over an AdS phase, and
during this period $\rho_\phi$ quickly redshifts away. Hereafter,
the field rapidly climbs up to the $\Lambda>0$ region, so that the
universe eventually settles down in the $\Lambda$CDM phase until
now. See also \cite{Piao:2004me,Li:2019ipk} for the potential with
multiple AdS vacua.

One has $\rho_\phi = \dot{\phi}^2/2+V(\phi)$ and
$P_\phi=\dot{\phi}^2/2-V(\phi)$, respectively. In the AdS phase,
$w=P_\phi/\rho_\phi>1$. When the EDE field rolls down to $V<0$, we
have $w>1$, so that $\rho_\phi$ redshifts very rapidly $\rho_\phi
\sim a^{-3(1+w)}$ (in Refs.\cite{Poulin:2018cxd,Agrawal:2019lmo,Smith:2019ihp}
the EDE dissipates less effectively by oscillation with
cycle-averaged $w<1$). This is crucial for getting a larger $H_0$,
since if $\rho_\phi$ is non-negligible around recombination, it
will interfere with the fit of $\Lambda$CDM to the CMB data \cite{Knox:2019rjx}.
Perturbations are also under control since a canonical scalar
field always has $c_s^2=1$.

As an example, we model Fig-\ref{Vads} as
\begin{equation}\label{V}
V(\phi)=\left\{\begin{aligned}
&V_0\left(\frac{\phi}{M_p}\right)^4-V_{ads},  &\frac{\phi}{M_p}<\left(\frac{V_{ads}}{V_0}\right)^{1/4}\\
&0, &\frac{\phi}{M_p}>\left(\frac{V_{ads}}{V_0}\right)^{1/4}
\end{aligned}\right.
\end{equation}
where $V_{ads}$ depicts the depth of AdS well. Here, the initial
value $\phi_i$ of $\phi$ should satisfy $|\phi_i|<M_P$, so that
the model is consistent with the swampland conjectures
\cite{Ooguri:2006in,Obied:2018sgi}. We also allow for a
cosmological constant $\Lambda\simeq (10^{-4}eV)^4>0$ (but not
included) in \eqref{V} to ensure that the universe eventually
settles down in the $\Lambda$CDM phase. It is possible to
replace the positive cosmological constant in $\Lambda$CDM with
quintessence, or some effective parameterization, e.g.\cite{Doran:2006kp}.
This may (but does not essentially) change the fit result. When
$V_{ads}=0$, \eqref{V} corresponds to a run-away potential, see
\cite{Lin:2019qug} for the relevant study.

Three new parameters $\{V_0,V_{ads},\phi_i\}$ are added to the
standard six parameters $\{\omega_b, \omega_{cdm},
H_0,\ln(10^{10}A_s),n_s,\tau_{reio}\}$ of $\Lambda$CDM, noting
initially $\dot{\phi}_i=0$. Instead of $\{V_0,V_{ads},\phi_i\}$,
we will adopt another set of parameters
$\{z_c,\omega_{scf},\alpha_{ads}\}$ with clearer physical
interpretation \cite{Agrawal:2019lmo,Poulin:2018cxd}. $z_c$ is the
redshift at which the EDE field starts rolling, which is defined
by $\partial^2_{\phi}V(\phi_c)=9H^2(z_c), \ \phi_c\equiv\phi(z_c)$
\cite{Marsh:2010wq}. $\omega_{scf}$ is the energy fraction of the
EDE field at $z_c$. $\alpha_{ads}$ is related to $V_{ads}$ by
$V_{ads} = \alpha_{ads}(\rho_m(z_c)+\rho_r(z_c))$.

In the code, one should search for the $\{V_0,V_{ads}\}$
corresponding to a given set of $\{z_c,\omega_{scf}\}$. A shooting
method is used to accomplish this. Exploiting the fact ${\dot \phi_i}=0$
before $z_c$, we get the initial guess by solving
$\partial^2_{\phi}V(\phi_i)=9H^2_c$ and
$V(\phi_i)=3\omega_{scf}M_p^2H^2_c$, where
$H_c^2\equiv(\rho_m(z_c)+\rho_r(z_c))/3M_p^2$. Then we search for
the exact scalar field parameters by iterately varying $V_{0,MC}$
and $\phi_{i,MC}$ and calculating the corresponding
$\omega_{scf,MC}$ with numeric integration.

\begin{figure}
    \includegraphics[width=\linewidth]{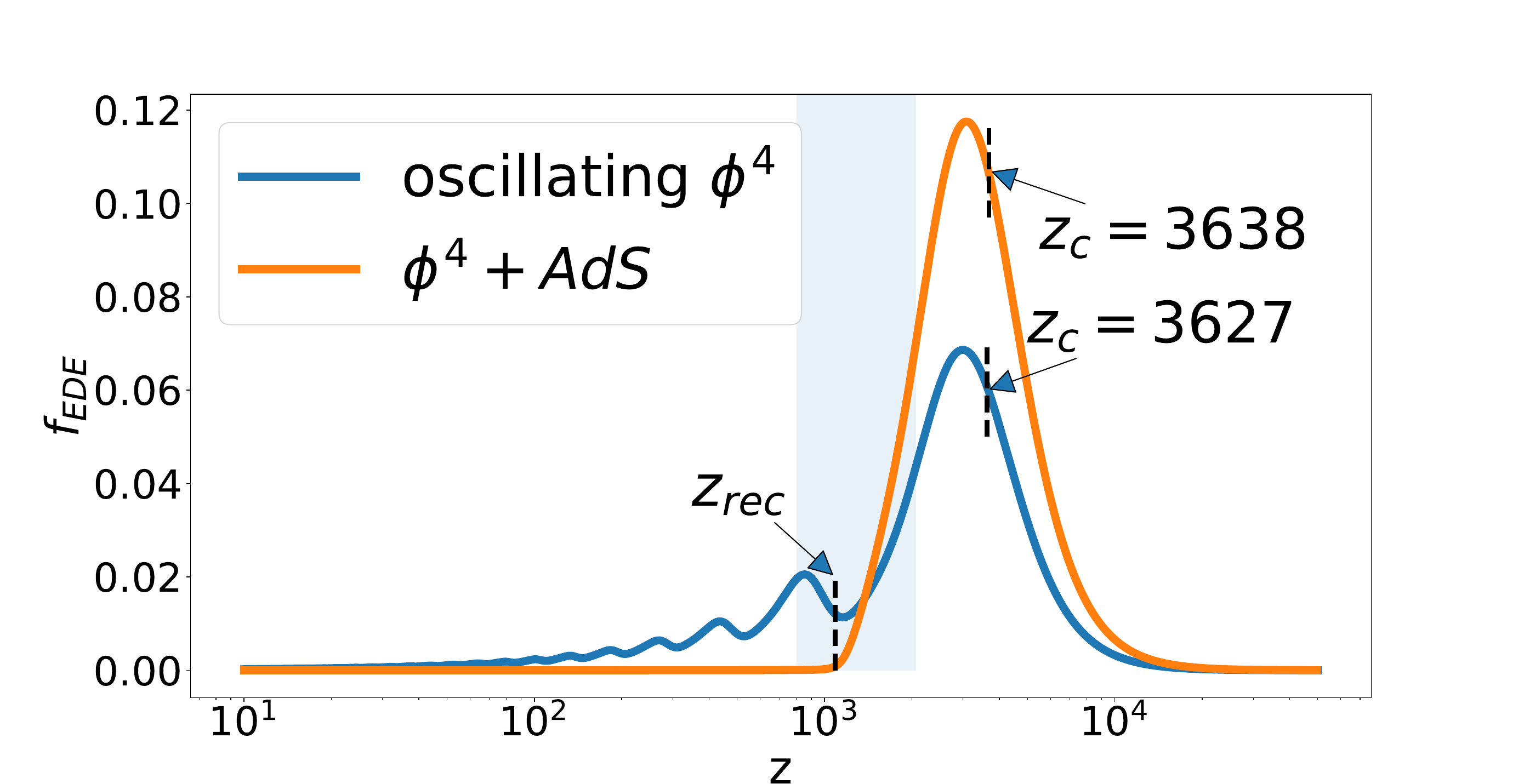}
    \caption{Energy fraction $f_{EDE}$ of EDE with respect to redshift
    $z$, plotted using the best-fit models. The scalar field energy
    density quickly redshifts away after the field starts rolling, so
    the recombination redshift $z_{rec}$ is nearly the same in both
    EDE models and the standard $\Lambda$CDM. Scalar field energy in
    the $\phi^4+AdS$ model redshifts much faster due to the AdS phase
    (shaded region, from $z=2063$ to $z=802$ in the best-fit model).}
    \label{f_EDE}
\end{figure}

We modified the MontePython-V3.2
\cite{Audren:2012wb,Brinckmann:2018cvx} and CLASS codes
\cite{Lesgourgues:2011re,Blas:2011rf} to perform a Markov chain
Monte Carlo (MCMC) analysis on the 6+2 parameter set $\{\omega_b,
\omega_{cdm},
H_0,\ln(10^{10}A_s),n_s,\tau_{reio},z_c,\omega_{scf}\}$. Our
datasets include Planck2018 high-$l$ and low-$l$ TT,EE,TE and
lensing likelihoods \cite{Aghanim:2019ame}. We follow the
convention used by Planck for the three neutrinos species. We use
BAO measurement from the CMASS and LOWZ of BOSS DR12
\cite{Alam:2016hwk} as well as low-z BAO measurements from 6dFGS
\cite{Beutler:2011hx} and MGS of SDSS \cite{Ross:2014qpa}. The
Pantheon \cite{Scolnic:2017caz} dataset with a single nuisance $M$, which includes
luminosity distance of 1048 SN Ia, is also included. We use the
latest result $H_0 = 74.03\pm 1.42$ km/s/Mpc from the SH0ES team
\cite{Riess:2019cxk} for local measurement.

One should also vary $\alpha_{ads}$ in the MCMC analysis. However,
variation of $\alpha_{ads}$ will drastically worsen
convergence of the chain. This is due to background integration
divergence whenever the field fails to climb up the potential.
As a consequence, the chain will head to low $\alpha_{ads}$ even
though better fit to data favors a higher one. In
principle, one should construct a compatible phase space measure
to account for this effect and adjust the step length accordingly.
Here, we will instead take a shortcut to simply fix $\alpha_{ads}$
to its best-fit value $\alpha_{ads}=3.79\times10^{-4}$, which is
enough for our purpose.

\begin{figure*}
    \includegraphics[width=\linewidth]{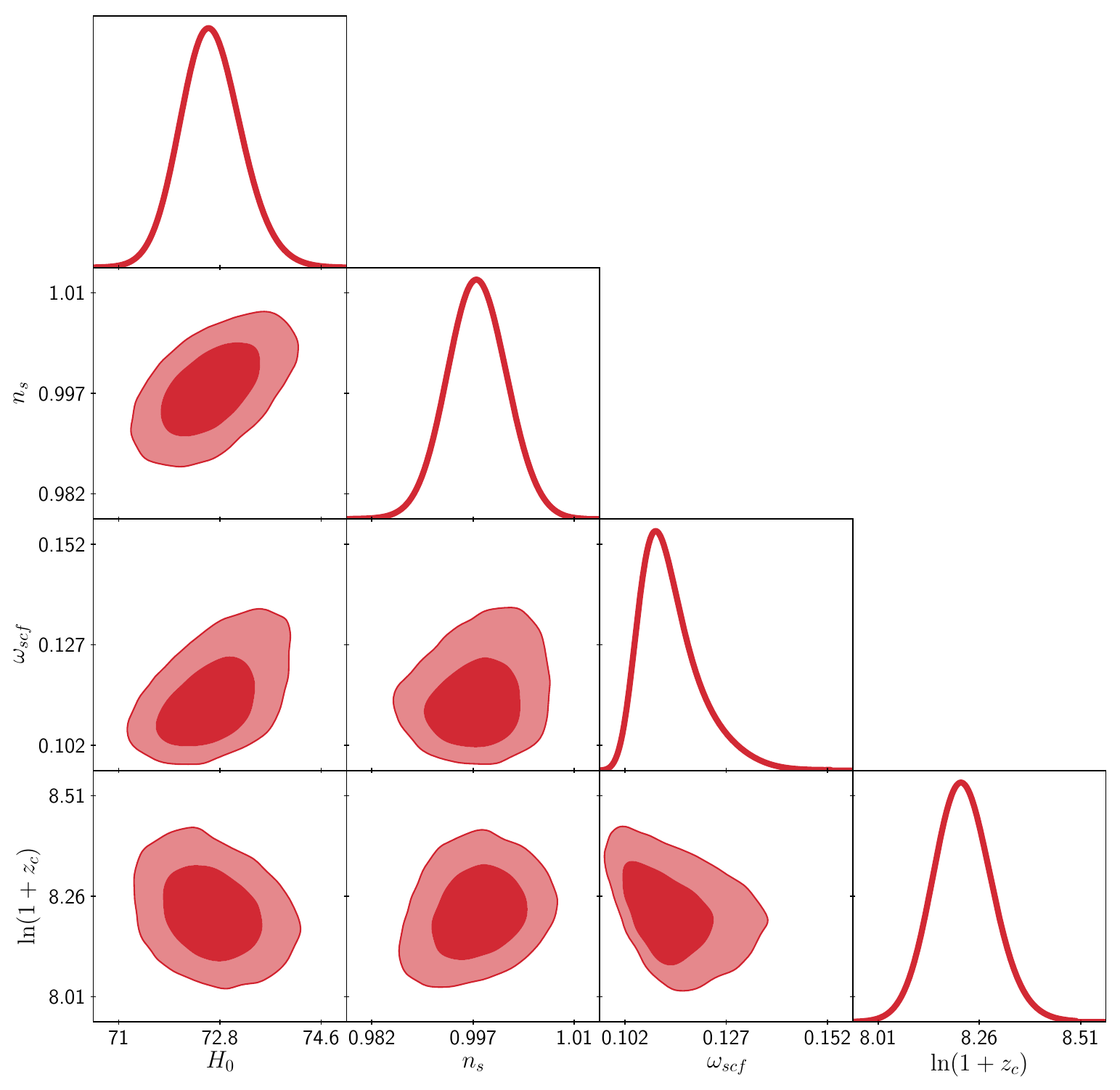}
    \caption{Marginalized posterior distributions of
$\{H_0,n_s,\omega_{scf},\ln(1+z_c)\}$. $H_0$ is correlated with
$n_s$ and $\omega_{scf}$, as explained in the text.}
    \label{triangle}
\end{figure*}

The marginalized posterior distributions of $\{H_0, n_s,
\omega_{scf}, \ln(1+z_c)\}$ are shown in Fig-\ref{triangle}. As
expected, the energy injection $\omega_{scf}$ is positively
correlated with $H_0$.
The mean and best-fit values of all model parameters are reported
in Table-\ref{cosmo_parameters}. Table-\ref{chi2} reports the
best-fit $\chi^2$ value of each individual experiment. The AdS model fits the CMB data slightly better than other models. We refer to our
model (\ref{V}) as the $\phi^4$+AdS model for simplicity. We also
include a $\Lambda$CDM model, an oscillating $\phi^4$ model
\cite{Agrawal:2019lmo} in which the EDE potential is $V(\phi)\sim
\phi^4$ and $\rho_\phi$ redshifts away when $\phi$ oscillates,
and a $\phi^4$+AdS model in the $\alpha_{ads}=0$ limit
(equivalently $V_{ads}=0$ in (\ref{V})) for comparison. In the
$\phi^4$+AdS model, the best-fit $H_0$ has been significantly
uplifted (as opposed to other models) to $72.74$ km/s/Mpc , in
agreement with the local measurements at 1$\sigma$ level.

\begin{figure}
\includegraphics[width=\linewidth]{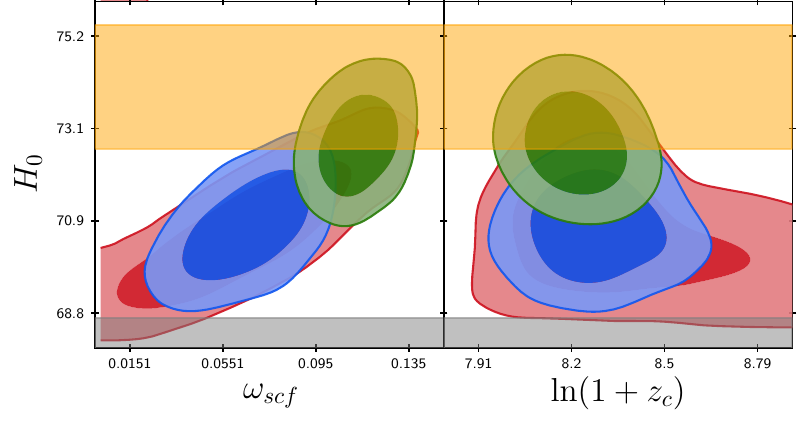}
\caption{The 1-sigma contour plot of $H_0$ versus $\omega_{scf}$
and $\ln(1+z_c)$. The models compared are oscillating $\phi^4$
(red), $\alpha_{ads}=0$ (blue) and
$\alpha_{ads}=3.79\times10^{-4}$ (green). The colored bands
represent the 1-sigma $H_0$ in $\Lambda$CDM (gray) and SH0ES
measurement (orange).} \label{comparison}
\end{figure}

In Fig-\ref{f_EDE}, we plot the evolution of
$f_{EDE}=\rho_\phi/\rho_{tot}$ with respect to the redshift for
the best-fit models.
The field thaws at $z=z_c$, quickly reaches the maximum of
$f_{EDE}$, and then $\rho_\phi$ rapidly redshifts away. Though
more energy is injected in the $\phi^4$+AdS model, $f_{EDE}$ at
recombination is far smaller, since the existence of an AdS phase
makes the dissipation of $\rho_\phi$ more effective. Our best-fit
model suggests that the recombination happened during the AdS
phase (or in AdS vacuum). To further illustrate the power of the
AdS phase, we compare the oscillating $\phi^4$ model, the
$\alpha_{ads}=0$ model and the $\alpha_{ads}=3.79\times10^{-4}$
model in Fig-\ref{comparison}. We see that though $V_{ads}$ only
takes up a quite small fraction [noting $V_{ads} =
\alpha_{ads}(\rho_m+\rho_r)$], its impact on $H_0$ is quite
remarkable.

\begin{figure}
    \subfigure{\label{delta:TT}}
    \includegraphics[width=\linewidth]{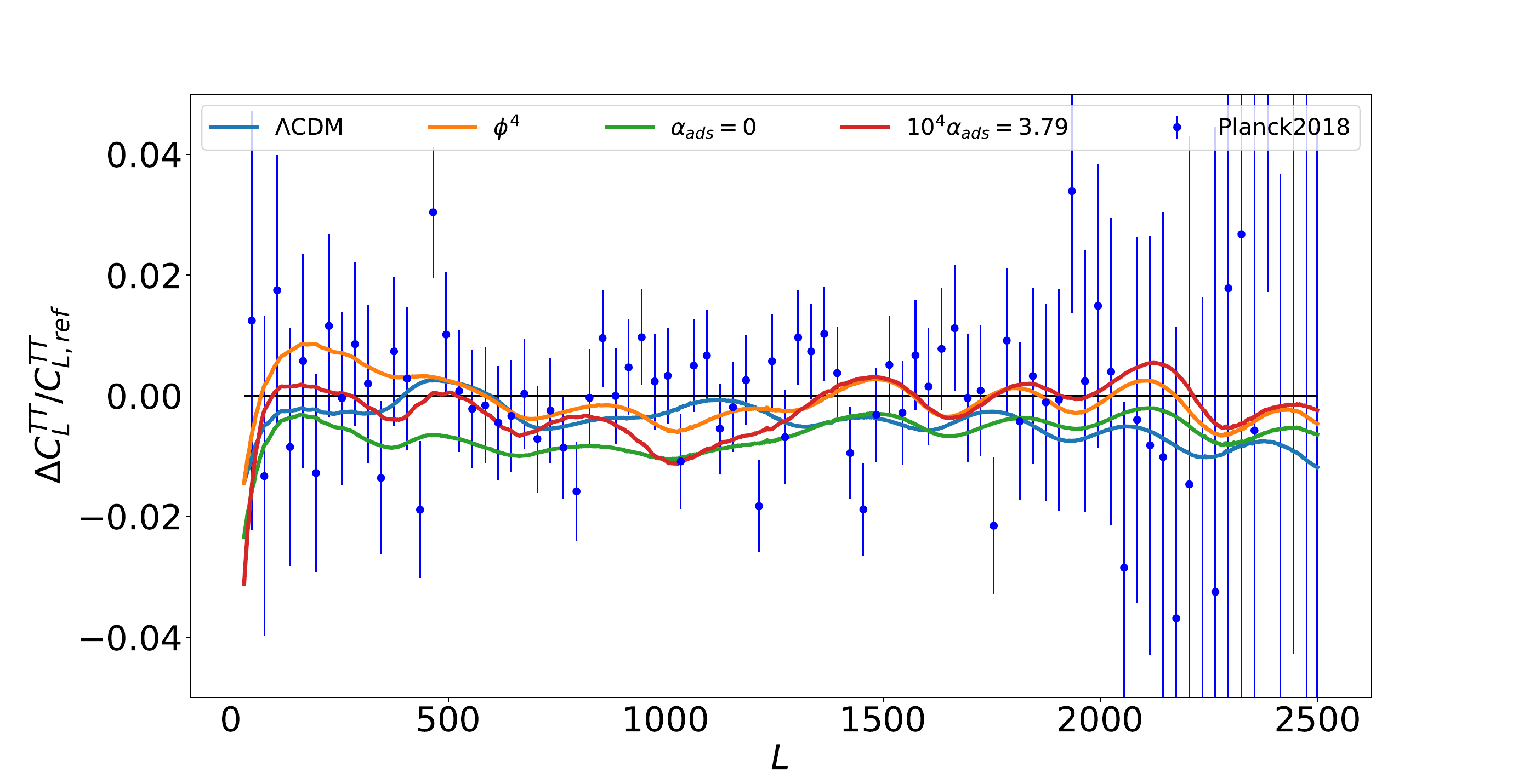}
    \subfigure{\label{delta:EE}}
    \includegraphics[width=\linewidth]{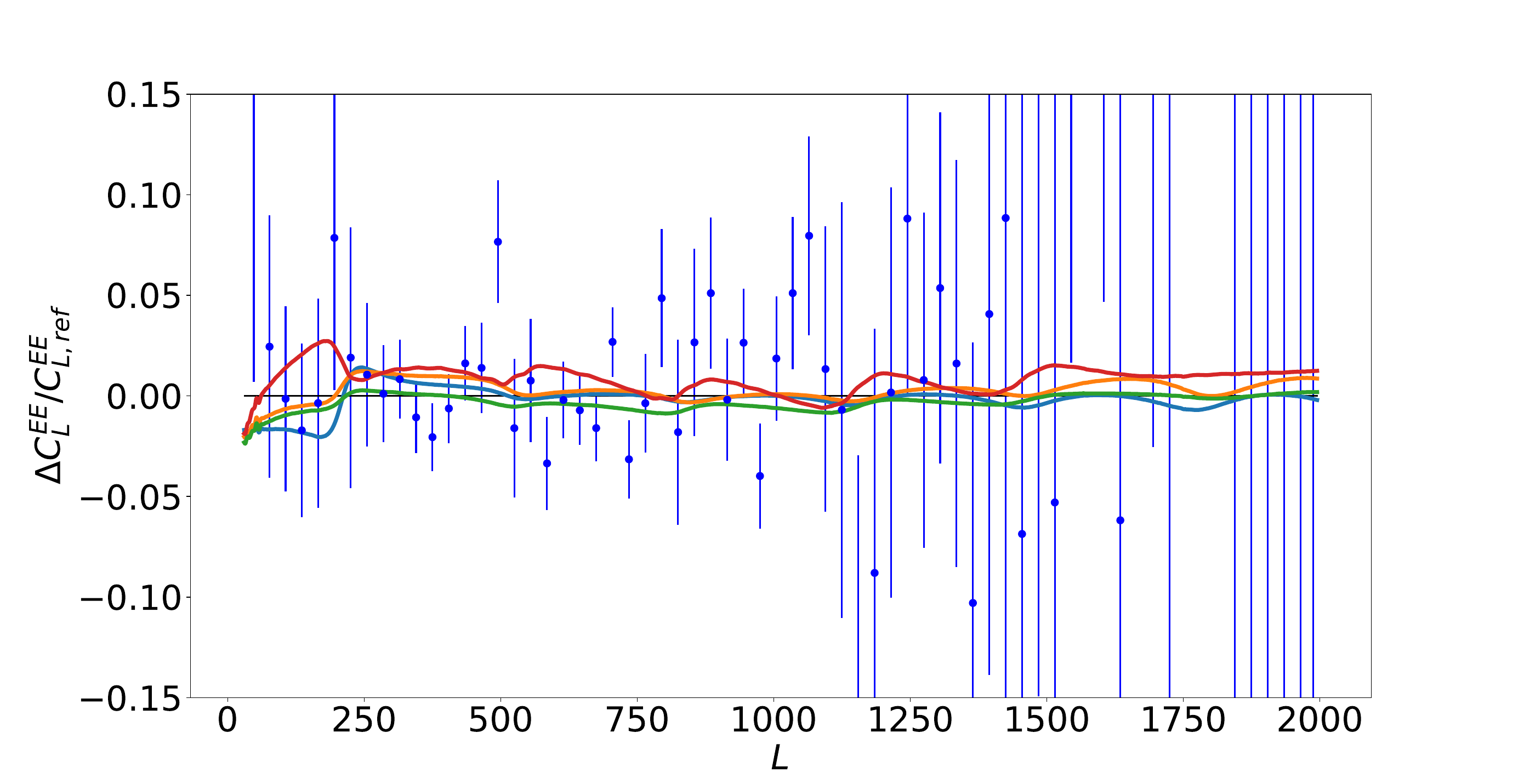}
\caption{Difference between various models fitted to full datasets
and a reference $\Lambda$CDM model obtained using only the
Planck2018 data. The upper panel is for the TT spectrum and the
lower one for the EE spectrum.}
    \label{delta}
\end{figure}

\begin{table*}
    \begin{tabular}{|c|c|c|c|c|}
    \hline
    \multirow{2}*{Parameter}& \multirow{2}*{$\Lambda$CDM}& \multirow{2}*{Oscillating $\phi^4$} & \multicolumn{2}{|c|}{$\phi^4$+AdS}  \\
    \cline{4-5}
    &&&$\alpha_{ads}=0$&$\alpha_{ads}=3.79\times10^{-4}$\\
    \hline
    $100~\omega_{b }$& $2.247(2.224)_{-0.014}^{+0.015}$ &$2.281(2.29)_{-0.02}^{+0.018}$&$2.301(2.289)_{-0.022}^{+0.02}$&$2.346(2.354)_{-0.016}^{+0.017}$ \\
    $\omega_{cdm }$& $0.1182(0.1183)_{-0.0013}^{+0.0008}$ &$0.1267(0.1256)_{-0.0048}^{+0.0038}$&$0.1275(0.1262)_{-0.0027}^{+0.0031}$ &$0.134(0.1322)_{-0.0021}^{+0.0019}$  \\
    $H_{0 }$ &$68.16(68.23)_{-0.4}^{+0.56}$ &$70.73(70.78)_{-1.3}^{+0.91}$ &$70.78(70.27)_{-0.71}^{+0.76}$& $72.64(72.74)_{-0.64}^{+0.57}$  \\
    $\ln(10^{10}A_{s })$ & $3.049(3.054)_{-0.016}^{+0.013}$ & $3.064(3.064)_{-0.018}^{+0.017}$ &$3.066(3.058)_{-0.017}^{+0.014}$& $3.077(3.074)_{-0.015}^{+0.015}$ \\
    $n_{s }$ & $0.9688(0.9696)_{-0.0042}^{+0.0039}$ &$0.9788(0.9798)_{-0.0064}^{+0.006}$&$0.9842(0.9805)_{-0.0057}^{+0.005}$ &$0.9976(0.9974)_{-0.0045}^{+0.0046}$  \\
    $\tau_{reio }$ & $0.0604(0.0636)_{-0.0075}^{+0.0066}$ &$0.0588(0.0575)_{-0.0084}^{+0.007}$&$0.0596(0.0573)_{-0.0084}^{+0.0086}$ &$0.0574(0.0598)_{-0.0078}^{+0.0075}$  \\
    $\omega_{scf}$& - &$0.0666_{-0.035}^{+0.029}$&$0.067(0.055)_{-0.015}^{+0.018}$ &$0.113(0.107)_{-0.009}^{+0.005}$ \\
    $\ln(1+z_{c})$& -
    & $8.347(8.197)_{-0.27}^{+0.11}$&$8.28(8.17)_{-0.13}^{+0.12}$ &$8.22(8.21)_{-0.079}^{+0.072}$\\
    \hline
    $100~\theta_{s }$ & $1.0422(1.0421)_{-0.0004}^{+0.0005}$ &$1.0415(1.0417)_{-0.0004}^{+0.0004}$&$1.0414(1.0415)_{-0.0005}^{+0.0006}$ &$1.0411(1.0411)_{-0.0003}^{+0.0003}$ \\
    $\sigma_{8 }$ & $0.8078(0.81)_{-0.0066}^{+0.0054}$ &$0.8368(0.8354)_{-0.011}^{+0.011}$&$0.835(0.8297)_{-0.0089}^{+0.0105}$ &$0.8571(0.8514)_{-0.0077}^{+0.0079}$  \\
    \hline
\end{tabular}
\caption{The mean values and 1$\sigma$ error of all cosmological
and model parameters. Best-fit values are given in the parenthesis.
The $\phi^4$+AdS models are labeled by their $\alpha_{ads}$
values. All models are obtained using the same datasets, nuisance
priors and precision settings.} \label{cosmo_parameters}
\end{table*}

\begin{table}
\begin{tabular}{|c|c|c|c|c|}
    \hline
    \multirow{2}*{Experiment}& \multirow{2}*{$\Lambda$CDM}& \multirow{2}*{Oscillating $\phi^4$} & \multicolumn{2}{|c|}{$\phi^4$+AdS}  \\
    \cline{4-5}
    &&&$\alpha_{ads}=0$&$10^4\alpha_{ads}=3.79$\\
    \hline
    $\chi^2_{CMB}$&2778.7&2782.1&2777&2776.5\\
    BAO low-$z$&2.2&2.1&1.8&2.2\\
    BAO high-$z$&1.8&1.9&1.9&2.1\\
    Pantheon&1026.9&1027.5&1026.9&1026.9\\
    SH0ES&15.4&4.9&7&0.8\\
    \hline
\end{tabular}
\caption{The best-fit $\chi^2$ per experiment.}
\label{chi2}
\end{table}

Here, the spectrum index $n_s$ and the amplitude $A_s$ of
primordial perturbations are larger than those in $\Lambda$CDM.
This is a common phenomenon in EDE scenarios, which has been also
observed in
Refs.\cite{Poulin:2018cxd,Agrawal:2019lmo,Smith:2019ihp}. In
particular, $n_s$ seems to be positively correlated with $H_0$,
see Fig-\ref{triangle}. It could be understood, at least
partially, by the integrated Sachs-Wolfe (ISW) effect
\cite{Sachs:1967}. The gravitational potential $\Phi$ can be
converted to the density perturbation $\delta$ through the Poison
equation $\nabla^2\Phi = 4\pi G\rho \delta$, thus the ISW contribution to the CMB angular
power spectrum is
\begin{equation}\label{Cl_1}(C_l)_{SW}
\propto \int_0^\infty
\frac{dk}{k}\mathcal{P}_\Phi(k)j^2_l(kD_A^*),
\end{equation} where
$\mathcal{P}_\Phi(k)$ is the power spectrum of primordial
perturbations in the Newtonian gauge. Considering
$\mathcal{P}_\Phi=A_s\left(\frac{k}{k_{pivot}}\right)^{n_s-1}$, we
can explicitly integrate out Eq.\eqref{Cl_1},
\begin{equation}\label{Cl_2}
(C_l)_{SW} \propto A_s\left(D_A^*k_{pivot}\right)^{1-n_s} f(n_s).
\end{equation}
In particular, $f(n_s)$ is monotonically increasing with respect
to $n_s$. In the EDE models, the reduction in $r_s^*$ causes a
reduction in $D^*_A$, and so the $\left(D_A^*k_{pivot}\right)^{1-n_s}$
term in Eq.\eqref{Cl_2} (for $n_s<1$). Since $C_l$ is fixed by the
CMB observation, $A_sf(n_s)$ must be larger accordingly,
eventually leading to larger $n_s$ and $A_s$.

We plot the difference $\Delta C_l/C_{l,ref}$ in Fig-\ref{delta},
where $\Delta C_l = C_{l,model} - C_{l,ref}$ and $C_{l,ref}$ is a
reference $\Lambda$CDM model obtained using only the Planck2018
dataset. The models compared are $\Lambda$CDM (fitted to the full
datasets), oscillating $\phi^4$ \cite{Agrawal:2019lmo} and
$\phi^4$+AdS with $\alpha_{ads}=0$ [equivalently $V_{ads}=0$ in
(\ref{V})] and $\alpha_{ads}=3.79\times10^{-4}$. It is observed
that as $l$ becomes large, compared with $\phi^4$ and $\Lambda$CDM
models, the TT spectrum in the $\phi^4$+AdS model will go upwards.
The near-future ground based CMB experiments
\cite{Ade:2018sbj,Abazajian:2016yjj,DiValentino:2016foa,Suzuki:2018cuy}
are expected to probe up to $l=5000$ with order of magnitude
improvement in precision at high $l$. The improved constraining
power from high $l$ will significantly help distinguishing
different models (especially those with different $n_s$ and
$A_s$). Another potentially observable signal is the EE
spectrum around $l\sim200$, which shows itself a bump for the
$\phi^4$+AdS model. The binned Planck residue at $l=165$ has $5\%$
relative uncertainty. The $l=200$ bump in the AdS model shows
roughly $5\%$ difference compared with the blue $\Lambda$CDM line thus it
is now invisible to the Planck observation. However, the upcoming
CMB-S4, for example, is expected to reduce uncertainty by $50\%$
over Planck and hence might be able to probe this signature.

In summary, we showed that the Hubble tension might be telling us
the existence of AdS vacua around recombination. Through studying
a phenomenological EDE model, we found that such an AdS phase
($|V_{ads}|\sim (0.1eV)^4$) can lift the CMB-inferred $H_0$ to
$H_0=72.64_{-0.64}^{+0.57}$ km/s/Mpc, within 1$\sigma$ range of
the local measurement \cite{Riess:2019cxk}, and significantly
alleviate the Hubble tension. A novelty of our result is that the
recombination happens in the AdS vacuum, which also makes unique
predictions accessible to near-future CMB experiments. It should
be pointed out that \eqref{V} is only applied for illustrating the
phenomenology of EDE with an AdS vacuum, and realistic potentials
may be more complex, which might fit better to data. Though the
model we consider is quite simplified, it highlights an unexpected
point that AdS vacua, ubiquitous in consistent UV-complete
theories, might also play a crucial role in our observable universe.

In order to have the EDE field start rolling at around
$z\sim 3000$ and be non-negligible, one must fine-tune the
potential such that $V_0\sim (eV)^4$. This fine-tuning guarantees
that the field thaws at the right time, otherwise it will not help
resolving the Hubble tension. It also should be pointed out that
on CMB probes, we only consider the Planck data. The $H_0$ tension is predominantly only present in Planck data and other CMB observatories do not see it in that magnitude. Our result might be weakened if there really are unknown
instrumental effects or systematics present in the Planck
experiment. It will be important to test our model against other
CMB datasets, e.g. WMAP, SPT and ACT, which will be studied
elsewhere. Multiprobe combination constraints on EDE models with
constant EDE fraction are also discussed in
Ref.\cite{Hollenstein:2009ph}.

The issues worth studying are as follows. A well-explored
conjecture is that AdS vacua is very likely to be accompanied by
an infinite tower of ultralight states
\cite{Ooguri:2006in,Klaewer:2016kiy,Lust:2019zwm} (this effect
also has recently been applied to the Hubble tension
\cite{Agrawal:2019dlm,Anchordoqui:2019amx}). It is quite
intriguing to explore whether these additional light states have
left any imprints on the last-scattering surface. It has recently
been proposed in \cite{Li:2019ipk} that a multistage inflation,
consisting of multiple inflationary phases separated by AdS vacua,
may survive the swampland conjectures. AdS-like potentials also
appear in nonsingular cosmological models \cite{Khoury:2001wf}.
Confronting these ideas with the hint from the Hubble tension that
the recombination era might happen in an AdS phase, it is quite
interesting to wonder if our universe actually has passed through
many phases with different AdS vacua.

\textbf{Acknowledgment} We thank Sunny Vagnozzi for valuable comments. G.Y. thanks Yu-Tong Wang for his help on
coding. Y.S.P. is supported by NSFC, No. 11575188 and No. 11690021.

\end{document}